\documentclass[12pt]{article}

\usepackage[pdftex,usenames,dvipsnames]{color}
\usepackage{graphics}
\usepackage{a4wide}
\usepackage{amsfonts}

\newcommand{\nit}{\noindent}

\newcommand{\np}{\newpage}
\newcommand{\dsp}{\displaystyle}
\newcommand{\vs}[1]{\vspace{#1 ex}}
\newcommand{\hs}[1]{\hspace{#1 em}}
\newcommand{\bflr}{\begin{flushright}}
\newcommand{\eflr}{\end{flushright}}
\newcommand{\bc}{\begin{center}}
\newcommand{\ec}{\end{center}}
\newcommand{\ben}{\begin{enumerate}}
\newcommand{\een}{\end{enumerate}}

\newcommand{\be}{\begin{equation}}
\newcommand{\ee}{\end{equation}}
\newcommand{\ba}{\begin{array}}
\newcommand{\ea}{\end{array}}
\newcommand{\ct}{\cite}
\newcommand{\bit}{\bibitem}

\newcommand{\ag}{\alpha}
\newcommand{\bg}{\beta}
\newcommand{\gam}{\gamma}
\newcommand{\del}{\delta}
\newcommand{\eps}{\epsilon}
\newcommand{\ve}{\varepsilon}

\newcommand{\thg}{\theta}
\newcommand{\kg}{\kappa}
\newcommand{\lb}{\lambda}
\newcommand{\sg}{\sigma}
\newcommand{\rg}{\rho}

\newcommand{\fg}{\phi}
\newcommand{\vf}{\varphi}
\newcommand{\og}{\omega}

\newcommand{\Fg}{\Phi}

\newcommand{\bftau}{\mbox{\boldmath{$\tau$}}}

\newcommand{\bfJ}{{\bf J}}

\newcommand{\bfW}{{\bf W}}

\newcommand{\cL}{{\cal L}}

\newcommand{\lh}{\left(}
\newcommand{\rh}{\right)}
\newcommand{\ld}{\left.}
\newcommand{\rd}{\right.}

\newcommand{\der}{\partial}

\begin{document}

\bflr
Nikhef/2015-21
\eflr
\vs{4}

\bc
{\large {\bf Completing the Standard Model}}\\
\vs{1}

{\large {\bf with right-handed neutrinos and $Z'$}}
\vs{8}

{\large J.W.\ van Holten}\footnote{e-mail: v.holten@nikhef.nl} 
\footnote{Nikhef lecture, May 11 2015} 
\vs{3} 

Nikhef, Amsterdam NL
\vs{3} 

\today
\ec
\vs{3}

\nit
{\small {\bf Abstract} \\
In this lecture I review the most relevant modifications of the Standard Model of particle physics 
that result from inclusion of right-handed neutrinos and a new neutral gauge boson $Z'$.}
\vs{6}

\nit
{\bf 1 Introduction} 
\vs{1}

\nit
The minimal Standard Model is the highly successful theory of quarks and leptons and their interactions
with vector and scalar fields. Most ingredients of the model have been developed in the period 1950-1975,
although the mass scales at which the various degrees of freedom come into play were unknown at the time. 
The last experimental key stone was put into place only in 2012 with the discovery of the Higgs boson by 
the LHC experiments at CERN. 

Two major modifications of the original model have since been made as a result of new experimental 
discoveries. First, a third family of quarks and leptons was discovered with large masses and short 
life times compared to those of the fermions already known. It was also established on the basis of 
$Z$-boson decay into light neutrinos that this third family presumably completes the set, at least of 
fermion families with lepton masses well below the $Z$-mass. The second major modification was 
the discovery of neutrino oscillations implying that neutrinos are massive, even though their masses 
must be extremely small compared to those of the charged leptons and quarks. 

A virtually inescapable consequence of non-vanishing neutrino masses is that the minimal Standard Model 
has to be extended with right-handed neutrinos even though they do not participate in the observed gauge 
interactions of the other quarks and leptons. This extension has more potential implications for the 
Standard Model. In particular it enlarges the set of symmetries that apply to quark- and lepton families;
thereby also new gauge interactions can come into the picture. 

In this lecture I discuss the minimal modification of the Standard Model by a new abelian gauge interaction, 
which does not require a modification of the family structure of the fermions. It does imply a new massive 
neutral vector boson, commonly denoted $Z'$. As we will see this modification can surprisingly also 
account for the observed tiny neutrino masses. This bonus comes for free and adds to the attraction of 
this modification. 
\vs{2}

\nit
{\bf 2 Overview of the Standard Model} 
\vs{1}

\nit
The Standard Model of particle physics is a relativistic quantum field theory of fermions with spin 
1/2 and bosons with spin 0 or 1 in units of $\hbar$. The fermions in the model describe three families 
of quarks and leptons, the vector and scalar bosons create the interactions between the various particles. 
Consistent quantum theories of interacting vector bosons are necessarily gauge-theories of massless 
quanta with only two polarization states; massive vector bosons with three polarization states arise by 
spontaneous breaking of the associated local gauge symmetry, in particular by interaction with a scalar 
field in a non-symmetric vacuum state: the Brout-Englert-Higgs (BEH) effect. 
\vs{1}

\bc
\begin{tabular}{|c|c|c|c|c|} \hline
particle & color & isospin & hypercharge & electric-charge  \\ 
   & multiplicity & $T_3$ & $Y$ & $Q  = Y/2 + T_3$ \\ \hline
 & & & & \\
$U_L$ & 3 & +1/2 & ~~1/3 & ~~2/3 \\
  & & & & \\
$D_L$ & 3 & $-1/2$ & ~~1/3 & $-1/3$ \\
 & & & & \\
$U_R$ & 3 & 0 & ~~4/3 & ~~2/3 \\
 & & & & \\
$D_R$ & 3 & 0 & $-2/3$ & $-1/3$ \\
 & & & & \\
$N_L$ & 1 & +1/2 & $-1$ & ~~0 \\
 & & & & \\
$E_L$ & 1 & $-1/2$ & $-1$ & $~-1$ \\
 & & & & \\
$N_R$ & 1 & 0 & ~~0 & ~~0 \\ 
 & & & & \\
$E_R$ & 1 & 0 & $-2$ & $~-1$ \\ 
& & & & \\ \hline
\end{tabular} 
\vs{1.5}

{\footnotesize Table 1: Gauge representations of a single family of quarks and leptons.}
\ec
\vs{2}

\nit
The gauge group of the Standard Model is $SU_c(3) \times SU_L(2) \times U_Y(1)$. Here the 
subscripts characterize the nature of the interactions: the $SU_c(3)$ gauge bosons are the gluons
mediating forces between color charges carried by quarks. The $SU_L(2)$ gauge fields interact 
only with left-handed quarks and leptons carrying isospin charges. And the $U_Y(1)$ fields 
create interactions between the hypercharges which are mostly non-zero but different for 
both left- and right-handed fermions. 

As a substantial part of the fermion interactions depend on the fermion chirality, and as chiral fermions 
are necessarily massless, also the masses of leptons and quarks arise by spontaneous symmetry 
breaking, such as Yukawa interactions with a BEH scalar field. In the minimal Standard Model 
this is the same scalar field which also generates the masses of vector bosons.

Table 1 summarizes some essential information on the gauge interactions of quarks and leptons.
The quarks are the only fermions carrying color charge; they transform according to the triplet 
representation of $SU_c(3)$. This is a vector-like interaction, acting on left- and right-handed $U$- 
and $D$-quarks in the same way. All left-handed fermions in the Standard Model carry isospin  
transforming as doublets w.r.t.\ $SU_L(2)$. The isospin $+1/2$ fermions in the doublets are the 
$U_L$ quarks and $N_L$ neutrinos, the isospin $-1/2$ fermions are the $D_L$ quarks and the 
$E_L$ (charged) leptons. 

All right-handed fermions are singlets of $SU_L(2)$, but they still carry a hypercharge to 
match the unbroken electric charges of their left-handed counterparts. The electric charge 
represents the coupling to the photon, the only combination of $SU_L(2)$- and $U_Y(1)$-gauge
fields continuing to represent massless particles after spontaneous symmetry breaking by the BEH 
scalar field. Electric charge is related to the isospin and hypercharge by
\be
Q = \frac{1}{2}\, Y + T_3.
\label{1.1}
\ee
In this way massive Dirac fermions can be formed with the same conserved electric charge for the 
left- and right-handed components.

The Standard Model also allows for the existence of anti-particles in right-handed doublets and 
left-handed singlets with opposite gauge charges. Right-handed neutrinos could be an exception 
to this rule, as in the minimal Standard Model the right-handed neutrinos are singlets w.r.t.\ all 
gauge symmetries. As such they do not interact directly with any of the gauge fields of 
$SU_c(3) \times SU_L(2) \times U_Y(1)$. Their only interaction is the Yukawa coupling to the 
left-handed leptons via the BEH field, yielding a Dirac mass for neutrinos. As neutrinos are known 
to be massive, it is highly probable that right-handed neutrinos exist and are involved in generating 
neutrino masses. 

In addition to local gauge symmetries the Standard Model also exhibits a number of rigid symmetries
implying conservation laws such as those for baryon- and lepton number; these amount to the counting
of the numbers of quarks minus anti-quarks and leptons minus anti-leptons, respectively.
\vs{2}

\nit
{\bf 3 Chiral, Majorana and Dirac fermions}
\vs{1}

\nit
An important characteristic of the Standard Model is the link between the chirality of fermions and 
their gauge charges. As only massless fermions can have a definite chirality, fermions can have 
non-vanishing masses only if some of these gauge charges are no longer conserved because the 
corresponding symmetries are spontaneously broken in the vacuum state. The relevance of these 
aspects of quantum field theory of fermions justifies a brief review of the various types of spinor fields 
used to describe fermions. 

Free fermions appear in quantum field theory as solutions of the Dirac equation\footnote{Here and 
in the following we take units such that $\hbar = c = 1$.}
\be
\lh i \gam \cdot p + m \rh \Psi(p) = 0,
\label{2.1}
\ee
where in general $\Psi$ is a complex 4-component spinor. This equation implies that the 4-momentum 
takes values on the mass shell:
\be
( - i \gam \cdot p + m)( i \gam \cdot p + m ) \Psi(p) = (p^2 + m^2) \Psi(p) = 0.
\label{2.2}
\ee
By taking the hermitean conjugate it follows that
\be
\bar{\Psi}(p) \lh i \gam \cdot p + m \rh = 0, \hs{2} \bar{\Psi} = \Psi^{\dagger} \gam_0.
\label{2.3}
\ee
Next we can apply charge conjugation. For fermions a charge conjugation matrix $C$ exists, which 
turns the transposed conjugate Dirac equation with 4-momentum $p$ into a ordinary Dirac equation 
with 4-momentum $-p$, as follows\footnote{A representation of the Dirac matrices and 
charge-conjugation matrix is found in appendix B.}:
\be
C \lh i \gam \cdot p + m \rh^T  \bar{\Psi}^T(p) = \lh - i \gam \cdot p + m \rh C \bar{\Psi}^T(p) = 0,
\label{2.4}
\ee
and we can define a charge-conjugate field satisfying the original Dirac equation: 
\be
\Psi^c(p) \equiv C \bar{\Psi}^T(-p) \hs{1} \Rightarrow \hs{1} \lh i \gam \cdot p + m \rh \Psi^c(p) = 0.
\label{2.5}
\ee
This result implies that solutions of the Dirac equation appear in pairs, corresponding to a particle state 
represented by $\Psi(p)$ and an anti-particle state represented by $\Psi^c(p)$. An exception is possible 
if the two solutions are identical:
\be
\Psi^c(p) = \Psi(p).
\label{2.6}
\ee
Spinor fields satisfying this condition are called Majorana spinors, and describe fermions which are 
their own anti-particle. 

Chirality is another constraint which can be imposed on spinors; specifically chiral spinors are 
eigenstates of $\gam_5$ labeled $L$ (left-handed) for a positive eigenvalue and $R$ (right-handed) 
for a negative eigenvalue :
\be
\gam_5 \Psi_L = \Psi_L, \hs{2} \gam_5 \Psi_R = - \Psi_R.
\label{2.7}
\ee
This eigenvalue constraint is Lorentz invariant, as is easy to verify:
\be
\left[ \gam_5, \sg_{\mu\nu} \right] = 0,
\label{2.8}
\ee
hence the eigenvalues of $\gam_5$ are the same in any Lorentz frame.
Chiral eigenstates can be constructed by applying chiral projection operators:
\be
\Psi_L \equiv \frac{1}{\sqrt{2}} \lh 1 + \gam_5 \rh \Psi, \hs{2}
\Psi_R \equiv \frac{1}{\sqrt{2}} \lh 1 - \gam_5 \rh \Psi. 
\label{2.9}
\ee
These projected states can satisfy the Dirac equation only in pairs:
\be
- i \gam \cdot p\, \Psi_L = m \Psi_R, \hs{2} - i \gam \cdot p\, \Psi_R = m \Psi_L.
\label{2.10}
\ee
A single left- or right-handed spinor cannot satisfy these equations unless $m = 0$. In particular 
this happens in the Standard Model, as left- and right-handed fermions carry different gauge
charges and transform differently under gauge transformations.

Clearly chiral spinors carry only half the number of degrees of freedom of full Dirac spinors. 
This is especially obvious in a representation of the Dirac matrices in which $\gam_5$ is 
diagonal\footnote{This is the representation actually provided in appendix B.}. Note however, that
the anti-particles of chiral fermions are massless fermions of the opposite chirality; mathematically
this follows as
\be
\gam_5 \Psi_L^{\,c} = - \Psi_L^{\,c}, \hs{2} \gam_5 \Psi_R^{\,c} = \Psi_R^{\,c},
\label{2.11}
\ee 
showing that the charge-conjugate spinor $\Psi_L^{\,c} \equiv (\Psi_L)^c$ is {\em right-handed} whilst 
$\Psi_R^{\,c} \equiv (\Psi_R)^c$ is {\em left-handed}.  The chiral nature of the Standard Model is 
particularly evident in the parity violation of $\bg$-decay, where always either left-handed neutrinos 
or right-handed anti-neutrinos are produced.

Like chiral spinors also Majorana spinors carry only half the physical degrees of freedom compared 
to the full Dirac field, as the constraint (\ref{2.6}) implies Majorana fermions to be their own anti-particles. 
However, as particles and anti-particles always have identical mass a Majorana fermion can possess 
mass without violating the Majorana condition. In contrast, Majorana fermions cannot have definite 
chirality, as chiral fermions are not their own anti-particles. In fact it can be verified that if $\Psi$ is a 
Majorana spinor (i.e., a spinor field invariant under charge conjugation), then $\gam_5 \Psi$ is an 
anti-Majorana spinor (i.e., a spinor field picking up a minus sign under charge conjugation):
\be
\Psi = \Psi^c \hs{1} \Rightarrow \hs{1} \lh \gam_5 \Psi \rh^c = - \gam_5 \Psi.
\label{2.12}
\ee 
Therefore the chiral projections (\ref{2.9}) are not invariant under charge conjugation and cannot be 
Majorana spinors. Indeed, under charge conjugation these projections are related by 
\be
\Psi = \Psi^c \hs{1} \Rightarrow \hs{1}  \Psi_L^{\,c} = \Psi_R, \hs{2} \Psi_R^{\,c} = \Psi_L,
\label{2.13}
\ee
in agreement with eqs.\ (\ref{2.11}). In physics terms these equations state that the right-handed polarization 
state of a Majorana fermion is the anti-state of the left-handed polarization state of the same particle.
\vs{2}

\nit
{\bf 4 Gauge couplings of fermions}
\vs{1}

\nit
In position space the Dirac equation for free fermions reads
\be
\lh \gam \cdot \der + m \rh \Psi(x) = 0.
\label{3.1}
\ee
Interactions with gauge fields are constructed by replacing the partial derivatives $\der_{\mu}$ by covariant 
derivatives $D_{\mu}$. As in the following we are not interested in strong interactions we ignore the coupling 
to QCD by putting all gluon fields equal to zero. Then the covariant derivatives w.r.t.\ the electro-weak
gauge group $SU_L(2) \times U_Y(1)$ take the form
\be
D \Psi_L = \lh \der - \frac{i}{2} \lh g_1 Y B + g_2 \bfW \cdot \bftau \rh \rh \Psi_L,  \hs{1}
D \Psi_R = \lh \der - \frac{i}{2}\, g_1 Y B \rh \Psi_R,
\label{3.2}
\ee
for $SU_L(2)$ doublets $\Psi_L$ and singlets $\Psi_R$ respectively. With these definitions the action for 
the massless gauge-coupled fermion fields is
\be
\ba{lll}
S_F & = & \dsp{ 
 \frac{i}{2}\, \int d^4x \left[ \sum_L \bar{\Psi}_L \gam \cdot D \Psi_L + \sum_R \bar{\Psi}_R \gam \cdot D \Psi_R \right] }\\
 & & \\
 & = & \dsp{ \frac{i}{2}\, \int d^4x \left[ \sum_L \bar{\Psi}_L \gam \cdot \der\Psi_L + 
 \sum_R \bar{\Psi}_R \gam \cdot \der \Psi_R - \frac{i}{2}\, \lh g_1 B \cdot J_{_Y} + g_2 \bfW \cdot \bfJ_{_L} \rh \right] }
\ea
\label{3.3}
\ee
the sums running over the doublets and singlets in table 1. The fermion currents coupling to the $U_Y(1)$
and $SU_L(2)$ gauge fields can be read off the explicit expressions for the covariant derivatives:
\be
\ba{lll}
J_{_Y \mu} & = & \dsp{ \frac{1}{3} \lh \bar{U}^a_L \gam_{\mu} U^a_L + \bar{D}^a_L \gam_{\mu} D^a_L \rh  
 + \frac{4}{3}\, \bar{U}^a_R \gam_{\mu} U^a_R - \frac{2}{3}\, \bar{D}^a_R \gam_{\mu} D^a_R }\\
 & & \\
 & & \dsp{ - \lh \bar{N}_L \gam_{\mu} N_L + \bar{E}_L \gam_{\mu} E_L \rh  - 2 \bar{E}_R \gam_{\mu} E_R, }\\
 & & \\
\bfJ_{_L \mu} & = & \dsp{ \bar{Q}^a_L\, \bftau\, \gam_{\mu} Q^a_L  + \bar{L}_L\, \bftau\, \gam_{\mu} L_L, }
\ea
\label{3.4}
\ee
where the summation over the quark color index $a = (1,2,3)$ has been made explicit. In the last line we 
have also introduced a short-hand notation for the doublets
\be
Q_L = \lh \ba{c} U_L \\ D_L \ea \rh, \hs{2} L_L = \lh \ba{c} N_L \\ E_L \ea \rh. 
\label{3.5}
\ee

\bc
\scalebox{0.13}{\includegraphics{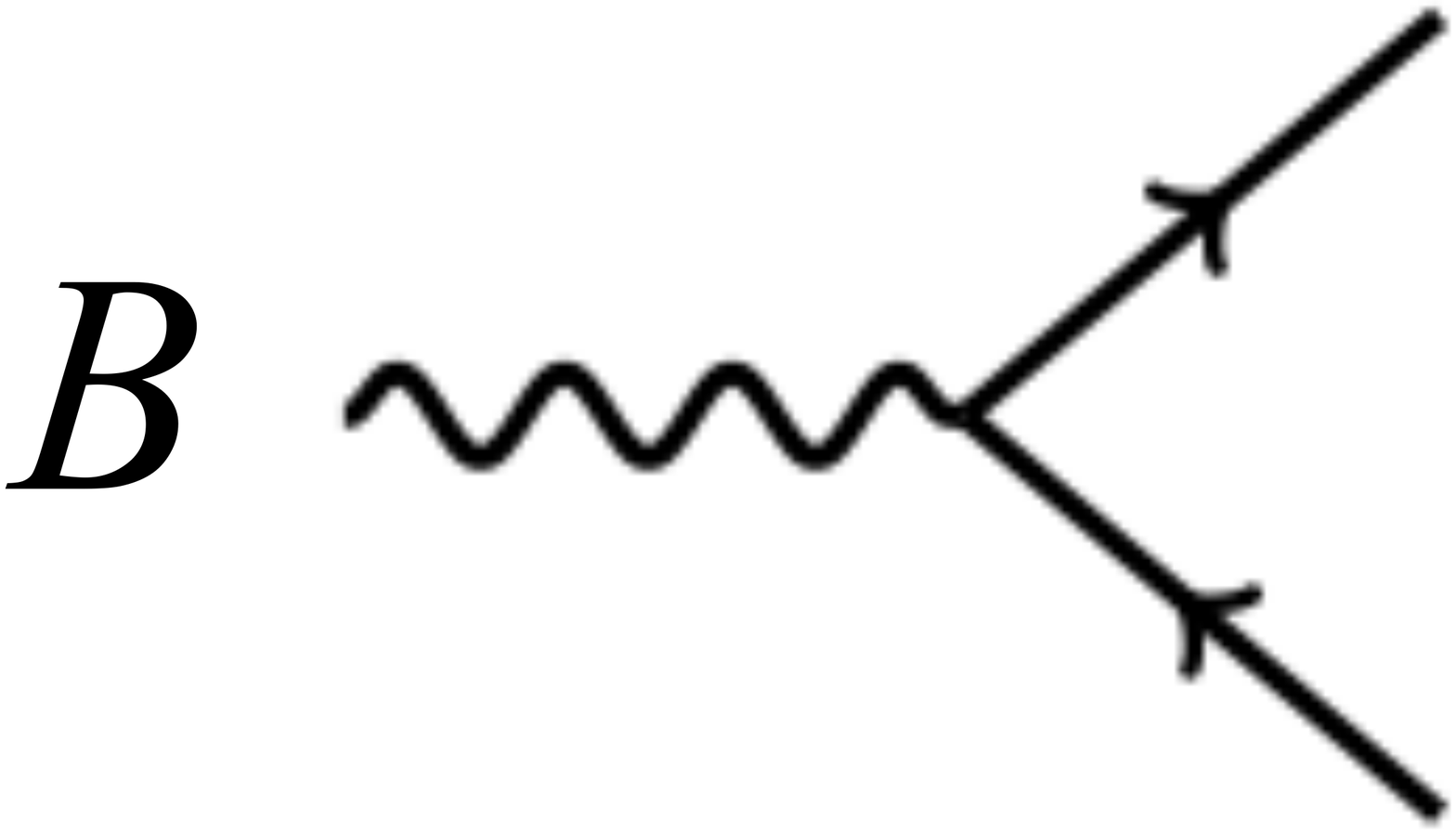}, \hs{1} \includegraphics{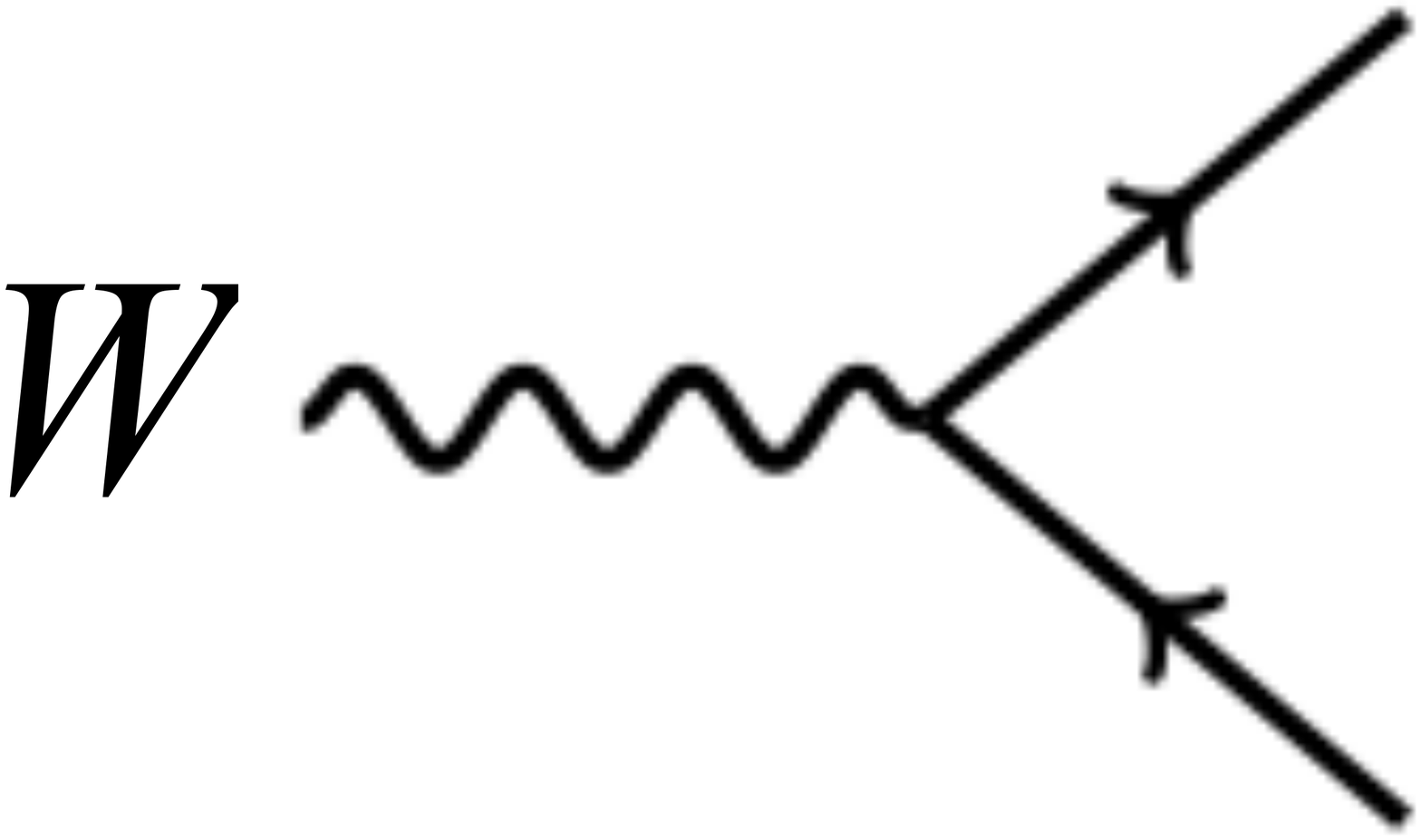}}
\vs{1}

{\footnotesize Fig.\ 1: Diagrammatic representation of fermion currents coupling to electro-weak gauge fields.}
\ec

\nit
{\bf 5 Chiral anomalies} 
\vs{1}

\nit
The classical field equation for a vector field like the hypercharge field is a variant of the Maxwell equations:
\be
\der_{\mu} F^{\mu\nu}(B) = J_{_Y}^{\mu}, \hs{2} \der_{\mu} J^{\mu}_{_Y} = 0.
\label{4.1}
\ee
The conservation law for the current follows directly from the field equation, but is a more general 
consequence of the $U_Y(1)$ gauge invariance of the theory. In momentum space it takes the 
form
\be
p \cdot J_{_Y}(p) = 0, 
\label{4.2}
\ee
where $p_{\mu}$ is the four-momentum of the gauge field coupling to the current. Diagrammatically 
this is the momentum flowing into the fermion current coupling to the gauge field in the diagrams of 
fig.\ 1. In principle such a condition should also hold in QFT when the $B$-field is coupled to a  
current appearing in some general Feynman diagram. In particular it should hold for a triangle 
diagram such as shown in fig.\ 2, where three vector fields interact by a fermion current running in
a closed loop. 
\vs{-2}

\bc
\scalebox{0.27}{\includegraphics{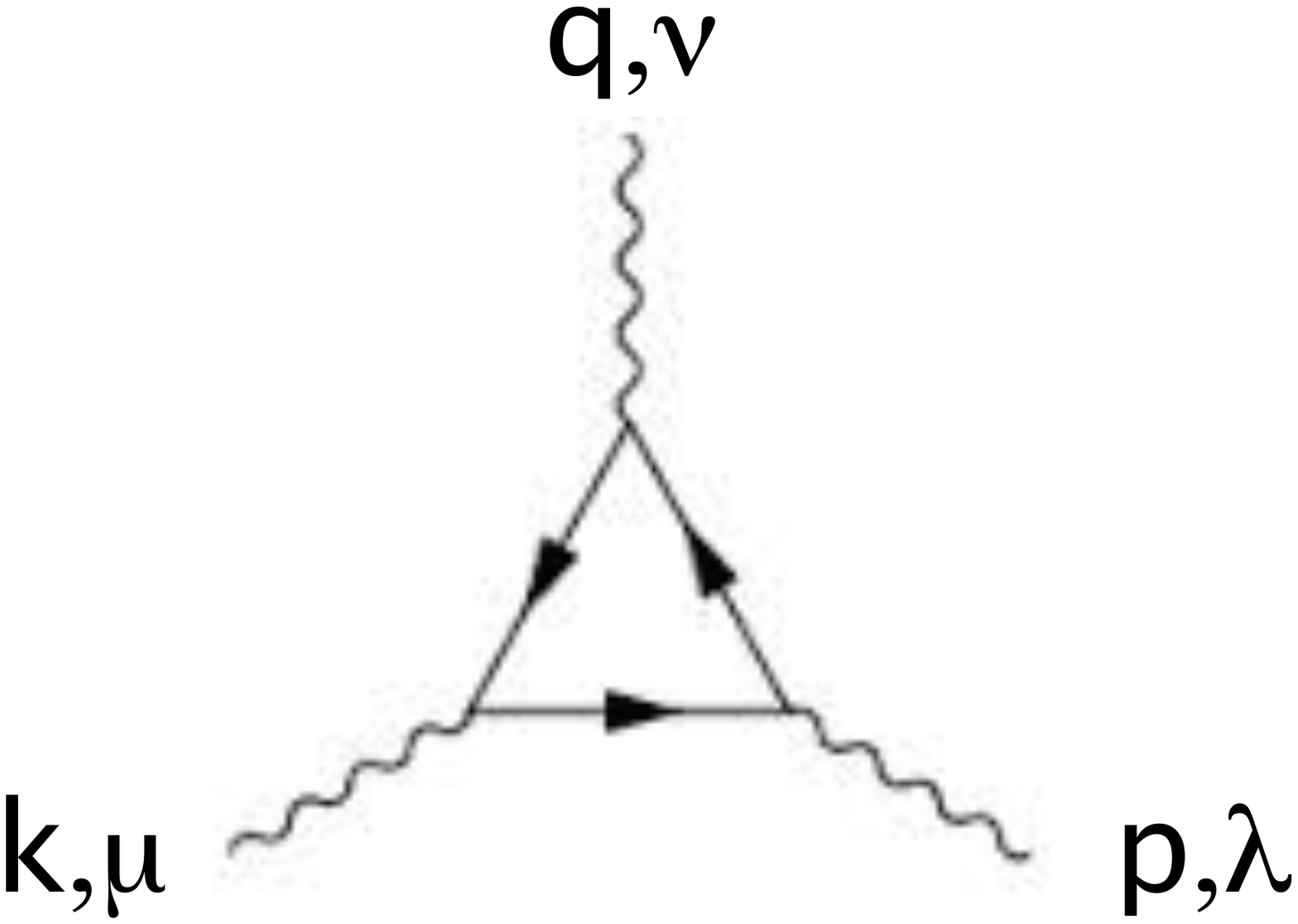}}
\vs{-3}

{\footnotesize Fig.\ 2: 3-vector interaction by fermion loop.}
\ec

\nit
If the four-momenta of the external vector bosons are $(k,q,p)$ and four-momentum is conserved, then the 
triangle diagram must be an expression $T_{\mu\nu\lb}(k,q)$ depending on the two independent momenta 
$k$ and $q$ with $p = -(k+q)$; the indices $\mu$, $\nu$ and $\lb$ label the vertices where the external 
vector fields couple to. Then based on our previous discussion this diagram should satisfy the identities
\be
k^{\mu} T_{\mu\nu\lb}(k,q) = q^{\nu} T_{\mu\nu\lb}(k,q) = (k+q)^{\lb} T_{\mu\nu\lb}(k,q) = 0.
\label{4.3}
\ee
Now it turns out that if the three vertices in the loop combined contain an odd number of $\gam_5$ matrices,
these conditions cannot be satisfied simultaneously for a single fermion in the loop\footnote{Briefly, to compute 
the diagram you need a regularisation scheme to avoid uninterpretable divergencies, and no regularisation 
scheme can maintain all three conditions at the same time.}. This creates a potential problem for the standard 
model, as the $(B, \bfW)$-fields couple to the chiral fermion currents (\ref{3.4}) and correspondingly the 
complete expression for a vertex in the loop is
\be
\frac{1}{\sqrt{2}}\, T_A \gam_{\mu} (1 \pm \gam_5),
\label{4.4}
\ee
where $T_A$ stands for the gauge charges like $(Y, \bftau)$. Hence the expression corresponding to the 
diagram will have contributions from terms with an odd number of $\gam_5$ matrices. However, there is a 
way out. The terms in the diagrams containing a $\gam_5$ (observing that $\gam_5^3 = \gam_5$) have 
coefficients
\be
\pm \mbox{Tr} \left[ \lh T_A T_B + T_B T_A \rh T_C \right],
\label{4.5}
\ee
where the sign is determined by the eigenvalue of $\gam_5$: +1 for left-handed fermions and $-1$ for right-handed
ones. Now the contributions for possible fermions in the loop have to be added, and as a result the contributions to 
the anomalous part of the 3-vector diagram cancel  each other if  
\be
\sum_l \mbox{Tr} \left[ \left\{ T_A^{(l)}, T_B^{(l)} \right\} T_C^{(l)} \right] 
 = \sum_{r} \mbox{Tr} \left[ \left\{ T_A^{(r)}, T_B^{(r)} \right\} T_C^{(r)} \right],
\label{4.6}
\ee
where $T_A^{(l,r)}$ are the charge matrices for left $(l)$ and right $(r)$ handed fermions. This is precisely what 
happens with the $SU_c(3) \times SU_L(2) \times U_Y(1)$ gauge charges in the Standard Model. As a relevant 
example, consider the hypercharges:
\be
\ba{l}
\dsp{ \sum_l Y^{(l)\,3}  =  3 \times 2 \times \lh \frac{1}{3} \rh^3 + 2 \times (-1)^3 = - \frac{16}{9}, }\\
 \\
\dsp{ \sum_r Y^{(r)\,3}  = 3 \times \lh \frac{4}{3} \rh^3 + 3 \times \lh - \frac{2}{3} \rh^3 + (-2)^3 = - \frac{16}{9}. }
\ea
\label{4.7}
\ee
Similar checks can be done for the other combinations of gauge charges. 
This guarantees that in the Standard Model the gauge symmetries and the corresponding conservation laws 
are observed including quantum corrections. 
\vs{2}

\nit
{\bf 6 Right-handed neutrinos and more gauge symmetries}
\vs{1}

\nit
In the discussions and calculations above the right-handed neutrinos do not play a role, as they possess
no gauge charges under $SU_c(3) \times SU_L(2) \times U_Y(1)$. It is therefore not very surprising that 
another anomaly-free $U(1)$ symmetry can be realized on the fermions in table 1 if the right-handed 
neutrinos are charged under this symmetry in an appropriate way. For example, one can define this 
symmetry to act on right-handed singlets only, as in table 2. 

\bc 
\begin{tabular}{|l|c|c|c|c|} \hline
particle & $U_R$ & $D_R$ & $N_R$ & $E_R$  \\ \hline
$R$-charge & 1 & $-1$ & 1 & $-1$ \\ \hline
\end{tabular}
\vs{2}

{\footnotesize Table 2: Anomaly free $U_R(1)$ charges of a family of quarks and leptons.}
\ec

\nit
More remarkable is, that one can define a continuous family of anomaly-free $U(1)$ symmetries by taking 
linear combinations of $U_Y(1)$ and $U_R(1)$: for any given fixed numbers $\ag$ and $\bg$ the charges
\be
X = \ag Y + \bg R,
\label{5.1}
\ee
specify a set of anomaly-free abelian transformations $U_X(1)$. An interesting example is $\ag = - \bg = 1$
in which case
\be
X = Y - R = B - L,
\label{5.2}
\ee
the difference between baryon and lepton number, as can be checked by inserting the $Y$ and $R$
charges in tables 1 and 2. 

The importance of an anomaly-free $U_X(1)$-symmetry is that it can be turned into a local gauge 
symmetry enriching the electro-weak sector of the Standard Model with an additional gauge field.
Actually there are much larger groups of symmetry transformations which can act anomaly-free on 
quarks and leptons, such as the grand-unified gauge groups $SU(5)$ or $SO(10)$ 
\ct{pati-salam1974,georgi-glashow1974,fritsch-minkowski1975}. However, these 
extensions embed the full Standard Model gauge group in a larger set of non-abelian symmetry 
transformations, many of which mix quarks and leptons. In contrast, the $U_X(1)$ symmetries 
defined above provide the only possible extensions of the Standard Model gauge group which 
commute with the original $SU_c(3) \times SU_L(2) \times U_Y(1)$ and leave the identity
of quarks and leptons untouched. It is also the minimal extension in which the right-handed 
neutrinos have gauge interactions with a vector field.
\vs{3}

\nit
{\bf 7 Electro-weak gauge theory of $SU_c(3) \times SU_L(2) \times U_Y(1) \times U_X(1)$.}
\vs{1}

\nit
Let us now see to what extent the Standard Model can be realistically completed with an additional 
$U_X(1)$ gauge symmetry \ct{buchmueller1991,leike1998,appelquist2002,langacker2009}. Certainly 
such a symmetry must be spontaneously broken and the new associated gauge field $C_{\mu}(x)$ 
must give rise to a massive gauge boson; otherwise we would observe long-range forces between 
quarks and leptons involving in particular also right-handed neutrinos. As no such interactions have been 
observed we exclude massless $C$-bosons. 

In principle there are many scenarios to generate a mass for the new vector boson. I will restrict myself 
here to the simplest option, interaction with a complex scalar field $\vf$ which is a singlet under 
$SU_L(2)$. Such a singlet scalar boson cannot couple directly to the $W$-bosons, therefore its 
vacuum expectation value does not contribute to the mass of the charged $W$ particles. However, 
a complex scalar can carry hypercharge; therefore the new scalar field can contribute to the mass of 
the $Z$-boson. In return, the usual BEH doublet field responsible for the $W$-mass can also carry 
$X$-charge; then it also contributes to the mass of the new vector particle, which will no longer 
be a purely $C$-quantum but a mixture of $C$ and $B$ usually denoted by $Z'$. Adding the new 
scalar singlet thus implies a more complicated mixing of neutral vector bosons in the electro-weak 
sector of the Standard Model. An important constraint on this mixing is that the photon should remain 
massless, as electromagnetism {\em is} an observed long-range force between electrically charged particles.
 
To interact with the $C$-field the complex singlet scalar must carry an $X$-charge, which we can choose to 
be the unit charge of reference; if it has an additional hypercharge $\eta$ the covariant derivative of the scalar 
field is
\be
D \vf = \lh \der - \frac{i}{2} \lh g_x C + g_1 \eta B \rh \rh \vf,
\label{6.1}
\ee
where $g_1$ and $g_x$ are the coupling constants of the hypercharge and $X$-charge field, respectively.
The scalar doublet of the Standard Model which breaks $SU_L(2) \times U_Y(1)$ and contains the
known Higgs particle couples to $W$- and $B$-fields in the standard way, and may in addition couple
to the $C$-field with charge $\xi$ :
\be
D \Fg = \lh \der - \frac{i}{2} \lh g_x\, \xi C + g_1 B + g_2 \bfW \cdot \bftau \rh \rh \Fg.
\label{6.2}
\ee
Now the scalars $\Fg$ and $\vf$ must get vacuum expectation values $(v_1, v_2)$  to generate 
masses for two neutral vector bosons. The most general invariant potential which can achieve this 
is
\be
V = \frac{\lb_1}{4} \lh |\Fg|^2 - v_1^2 \rh^2 + \frac{\lb_2}{4} \lh |\vf|^2 - v_2^2 \rh^2
 + \frac{\lb_m}{4} \lh |\vf|^2 - v_2^2 \rh \lh |\Fg|^2 - v_1^2 \rh.
\label{6.3}
\ee
Minimizing this potential in the unitary gauge leads to the physical field parametrization
\be
\Fg = \lh \ba{c} 0 \\ v_1 + h/\sqrt{2} \ea \rh, \hs{2} \vf = v_2 + \frac{a}{\sqrt{2}},
\label{6.4}
\ee
where $h$ and $a$ represent the Higgs fields left over after the vector bosons have become massive.
Now let us write out the covariant kinetic terms of the scalar fields in the action in this parametrization:
\be
\ba{lll}
- |D\Fg|^2 - |D\fg|^2 & = & \dsp{ - \frac{g_2^2 v_1^2}{2}\, W^+ W^-  
 - \frac{v_1^2}{4} \lh g_x\, \xi C + g_1 B - g_2 W_3 \rh^2 }\\
 & & \\
 & & \dsp{ - \frac{v_2^2}{4} \lh g_x C + g_1 \eta B \rh^2 + \mbox{(terms involving the Higgs fields $(h,a)$}. }
\ea
\label{6.5}
\ee
One can observe immediately that the masses of the charged $W$-bosons are given by the standard 
result
\be
m_W^2 = \frac{g_2^2 v_1^2}{2}.
\label{6.6}
\ee
In addition two combinations of neutral vector bosons become massive: 
\be
Z = \frac{g_2 W_3 - g_x\, \xi C - g_1 B}{\sqrt{g_1^2 + g_2^2 + g_x^2 \xi^2}}, \hs{2}
Z' = \frac{g_x C + g_1 \eta B}{\sqrt{g_x^2 + g_1^2 \eta^2}}, 
\label{6.r7}
\ee
with masses
\be
m_Z^2 = \frac{v_1^2}{2} \lh g_1^2 + g_2^2 + g_x^2 \xi^2 \rh, \hs{2}
m_{Z'}^2 = \frac{v_2^2}{2} \lh g_x^2 + g_1^2 \eta^2 \rh.
\label{6.r8}
\ee
The third linear combination of $(W_3, B, C)$ is the photon $\gam$ and must remain massless. 
The price to be paid for a massless photon is, that the coupling parameters $\xi$ and $\eta$
are related:
\be
\frac{\eta}{\xi} = \frac{g_x^2}{g_1^2} \hs{1} \Rightarrow \hs{1} 
 Z' = \frac{g_1 C + g_x \xi B}{\sqrt{g_1^2 + g_x^2 \xi^2}}.
\label{6.r9}
\ee
Upon implementing this constraint, a convenient parametrization of the physical states is in terms of two 
mixing angles $(\thg_w, \del)$, such that
\be
\ba{l}
\gam = W_3 \sin \thg_w + \lh B \cos \del - C \sin \del \rh \cos \thg_w, \\
 \\
Z = W_3 \cos \thg_w - \lh B \cos \del - C \sin \del \rh \sin \thg_w, \\
 \\
Z' = B \sin \del + C \cos \del.
\ea
\label{6.r10}
\ee
This is equivalent to the combinations (\ref{6.r7}) provided 
\be
\ba{ll}
\dsp{ \cos \del = \frac{g_1}{\sqrt{ g_1^2 + g_x^2 \xi^2}}, }& 
\dsp{ \sin \del = \frac{g_x \xi}{\sqrt{ g_1^2 + g_x^2 \xi^2}}, }\\
 & \\
\dsp{ \cos \thg_w = \frac{g_2}{\sqrt{g_1^2 + g_2^2 + g_x^2\, \xi^2}}, }&
\dsp{ \sin \thg_w = \sqrt{\frac{g_1^2 + g_x^2 \xi^2}{g_1^2 + g_2^2 + g_x^2\, \xi^2}}. }
\ea
\label{6.r11}
\ee
The resulting non-zero masses for the vector bosons $(Z, Z')$ are 
\be
m_Z^2 = \frac{g_2^2 v_1^2}{2 \cos^2 \thg_w}, \hs{2}
m_{Z'}^2 = \frac{g_x^2 v_2^2}{2 \cos^2 \del}.  
\label{6.10}
\ee
The equation for $m_Z$ can be converted to an equivalent expression for a $\rg$-parameter defined in terms
of the redefined mixing angle $\thg_w$ of electroweak interactions:
\be
\rg =  \frac{m_W^2}{m_Z^2 \cos^2 \thg_w} = 1,
\label{6.7}
\ee
implied by (\ref{6.r9}) and signifying a massless photon.
\vs{2}

\nit
{\bf 8 Fermions and Yukawa couplings}
\vs{1}

\nit
In the construction of the interactions of vector and scalar bosons the $X$-charge of the $\vf$-boson 
has been set equal to one by fiat (as it cannot vanish), whilst in addition it may possess the hypercharge 
$\eta$.  The $X$-charges of other fields, such as the scalar doublet $\Fg$ and the fermions are defined
relative to this unit. The covariant derivatives (\ref{3.2}) of the fermions are then extended to
\be
D \Psi_L = \lh \der - \frac{i}{2} \lh g_x X C + g_1 Y B + g_2 \bfW \cdot \bftau \rh \rh \Psi_L,  \hs{0.5}
D \Psi_R = \lh \der - \frac{i}{2} \lh g_x X C + g_1 Y B \rh \rh \Psi_R,
\label{7.r0}
\ee
Recall that according to eq.\ (\ref{5.1}) the $X$-charges of the fermions can be 
decomposed as 
\[
X = \ag Y + \bg R.
\]
The possible sets of fermion $U(1)$-charges are therefore as summarized in table 3. 
\vs{1}

\bc
\begin{tabular}{|c|c|c|c|c|c|c|} \hline
   & $Q_L$ & $U_R$ & $D_R$ & $L_L$ & $N_R$ & $E_R$ \\ \hline
$X$ & $\ag/3$ & $4\ag/3 + \bg$ & $-2\ag/3 - \bg$ & $-\ag$ & $\bg$ & $-2\ag - \bg$ \\
$Y$ & 1/3 & 4/3 & $-2/3$ & $-1$ & 0 & $-2$ \\ \hline
\end{tabular} 
\vs{2}

{\footnotesize Table 3: $X$- and $Y$-charges for a family of Standard Model fermions.}
\ec

\nit
In this parametrization $\bg \neq 0$, otherwise the $X$ charge reduces to the hypercharge $Y$
and there is no new independent gauged $U_X(1)$ symmetry acting on the fermions. Now in the
minimal Standard Model the physical fermions acquire a mass via Yukawa couplings to the same 
scalar doublet BEH field generating the mass of the $W$-bosons:
\be
\cL_{Yuk} = - \frac{i}{2}\, \eps_{ab} \Fg^a \left[ f_U \bar{U}_R Q^b_L + f_N \bar{N}_R L^b_L \right]  
                 - \frac{i}{2}\, \Fg^*_a \left[ f_D \bar{D}_R Q^a_L + f_E \bar{E}_R L^a_L \right] + c.c.\
\label{7.r1}
\ee
For the sake of simplicity I have not  made explicit family mixing in my notation, but it can be included
in a straightforward manner if necessary \ct{cabibbo1963,kobayashi-maskawa1973}. In the present 
extension of the Standard Model we need to make sure that these couplings respect the $U_X(1)$
 gauge symmetry. In table 4 I have collected the $X$-charges of the relevant bilinear fermion 
 combinations following from table 3:              
\vs{1}

\bc
\begin{tabular}{|c|c|c|c|c|} \hline
bilinear & $\bar{U}_R Q_L$ & $\bar{D}_R Q_L$ & $\bar{N}_R L_L$ & $\bar{E}_R L_L$ \\ \hline
$X$ & $-(\ag + \bg)$ & $\ag + \bg$ & $-(\ag+ \bg)$ & $\ag + \bg$ \\ \hline
\end{tabular} 
\vs{2}

{\footnotesize Table 4: $X$-charges of fermion bilinears in the standard BEH-doublet couplings.}
\ec

\nit
These results show, that all Yukawa couplings (\ref{7.r0}) are $X$-invariant when the charge $\xi$ 
of the scalar doublet $\Fg$ is taken to be 
\be
\xi = \ag + \bg.
\label{7.r2}
\ee
Therefore the $U_X(1)$ gauge symmetry can act on the fermions and the BEH doublet without spoiling the 
successes of the minimal Standard Model. 

In the presence of the singlet scalar $\vf$ there may also arise new Yukawa couplings of the type
\be
\vf\, \bar{\Theta}_R \Psi_L + c.c.,
\label{7.1}
\ee
if  there exist fermion bilinears $\bar{\Theta}_R \Psi_L$ which are singlets under $SU_c(3) \times S_LU(2)$ 
but not under $U_Y(1) \times U_X(1)$. Invariance under $SU_c(3)$ excludes quark bilinears. 
However, in the lepton-sector such invariant bilinears can be constructed indeed. $SU_L(2)$-invariance 
and chirality dictates that these are Majorana-type bilinears:
\be
L_L^T C \tau_2 L_L, \hs{1} E_R^T C E_R, \hs{1} N_R^T C N_R.
\label{7.2}
\ee
However, the first two bilinears would generate a Majorana mass for charged leptons, which is impossible 
as it breaks $U_{em}(1)$ and violates electric charge conservation. Therefore only the Majorana term for 
the right-handed neutrinos is phenomenologically viable. The unique allowed invariant Yukawa coupling 
of Standard Model fermions to the complex scalar singlet is of the form
\be
\frac{1}{2} \lh f_x\, \vf^*\, N_R^T C N_R + f^*_x\, \vf\, \bar{N}_R C \bar{N}_R^T \rh,
\label{7.3}
\ee
with $f_x$ the Yukawa coupling constant for the right-handed singlet neutrinos; it is $U_X(1)$-invariant 
only if 
\be
\bg = \frac{1}{2} \hs{1} \Rightarrow \hs{1} \ag = \xi - \frac{1}{2}.
\label{7.4}
\ee
A side remark: the choice $\bg = -1/2$ and interchange of $\vf \leftrightarrow \vf^*$ 
does not produce a new result, as in the scalar potential only $|\vf|^2$ appears and the vector-boson mass 
matrix is unchanged by the replacement of $\vf$ by its complex conjugate. Plugging the values (\ref{7.4}) 
into the entries in table 3 gives finally gives the fermion $X$-charges to have the value \vs{2}
\be
X = \xi Y - \bg (B - L). 
\label{7.r3}
\ee
Clearly if $\xi = 0$ the $X$-charge is identical with $B-L$, up to scaling which can be absorbed in a redefinition
of the gauge coupling $g_x$. In this case the BEH-doublet is invariant under the $U_{B-L}(1)$ symmetry and 
as $\sin \del = 0$  the $Z'$-boson is purely a massive $C$-boson. 

Finally note that irrespective of the choice of $\xi$ the Majorana-Yukawa coupling (\ref{7.3}) of the right-handed 
neutrino preserves the rigid $(B-L)$-symmetry separately if we assign the scalar singlet $\vf$ a 
$(B-L)$-charge 2. The complete lagrangean density of the $U_X(1)$-completed electro-weak Standard Model 
can be found in appendix A for reference.
\vs{2}

\nit
{\bf 9 Neutrino masses} 
\vs{1}

\nit
For the parameter choice (\ref{7.4}) the right-handed neutrinos can get a Majorana mass by the 
BEH-mechanism for the scalar singlet $\vf$. However, the coupling with the left-handed neutrinos
via the standard BEH-doublet with Yukawa coupling $f_N$ also generates Dirac masses. Therefore 
the total neutrino mass matrix (ignoring family mixing) is generated by a non-trivial combination of 
the two BEH-mechanisms, which reproduces the well-known see-saw mechanism for neutrino 
masses \ct{minkowski1977,gellmann1979}.

The analysis starts by switching to a basis of Majorana spinors for the neutrinos as explained in sect.\ 3.
Thus we introduce two Majorana spinors $(N_s, N_d)$ defined by
\be
N_R = \frac{1}{\sqrt{2}} \lh 1 - \gam_5 \rh N_s, \hs{2}
N_L = \frac{1}{\sqrt{2}} \lh 1 + \gam_5 \rh N_d.
\label{8.1}
\ee
In terms of these the Yukawa couplings for the neutrinos generate mass terms
\be
\frac{m_D}{2} \lh \bar{N}_s N_d + \bar{N}_d N_s \rh  + \frac{M}{2}\, \bar{N}_s N_s
 = \frac{1}{2} \lh \bar{N}_d, \bar{N}_s \rh \lh \ba{cc} 0 & m_D \\
                                                                                 m_D & M \ea \rh \lh \ba{c} N_d \\ N_s \ea \rh,
\label{8.2}
\ee
with
\be
m_D = f_N v_1, \hs{2} M = f_x v_2.
\label{8.3}
\ee
This mass matrix can easily be diagonalized to give the eigenvalues
\be
m_{\pm} = \frac{1}{2} \lh M \pm \sqrt{ M^2 + 4 m_D^2} \rh,
\label{8.4}
\ee
and eigenstates
\be
\ba{l}
N_- = N_d \cos \thg_{\nu} - N_s \sin \thg_{\nu}, \hs{1} N_+ = N_d \sin \thg_{\nu} + N_s \cos \thg_{\nu}, \\
 \\
\dsp{ \tan 2 \thg_{\nu} = \frac{2 m_D}{M}. }
\ea
\label{8.5}
\ee
In the limit $m_D \ll M$ the mass eigenvalues reduce in good approximation to
\be
\ba{l}
\dsp{ m _- = \frac{m_D^2}{M} \hs{1} \Rightarrow \hs{1} 
 \frac{m_-}{m_Z} = \sqrt{2}\; \frac{\cos^2 \thg_w}{\cos \del}\, \frac{g_x f_N^2}{g_2^2 f_x}\, \frac{m_Z}{m_{Z'}}, }\\
 \\
\dsp{ m_+ = M \hs{1} \Rightarrow \hs{1} \frac{m_+}{m_{Z'}} = \sqrt{2}\; \frac{f_x}{g_x}, }
\ea
\label{8.6}
\ee
whilst $N_- \approx N_d$, $N_+ \approx N_s$. From  neutrino oscillation experiments involving the light neutrinos 
it is infered that the left-hand side of the first equation satisfies $m_-/m_Z \sim 10^{-12}$. However, a neutrino 
Yukawa coupling $f_N$ of the same order of magnitude as that of the electron implies 
$f_N^2 \sim f_E^2 \sim 10^{-11}$. In such a scenario a $Z'$ mass in the TeV region is possible if $\del$ is 
sufficiently small ($\cos \del \sim1 $).
\vs{2}

\nit
{\bf 10 Scalar masses} 
\vs{1}

\nit
I conclude this analysis with the computation of the scalar masses for the physical Higgs particles
\ct{buchmueller1991}. The starting point is the potential (\ref{6.3}):
\[
V = \frac{\lb_1}{4} \lh |\Fg|^2 - v_1^2 \rh^2 + \frac{\lb_2}{4} \lh |\vf|^2 - v_2^2 \rh^2
 + \frac{\lb_m}{4} \lh |\vf|^2 - v_2^2 \rh \lh |\Fg|^2 - v_1^2 \rh.
\]
In this potential we substitute the Ansatz (\ref{6.4}) representing the unitary gauge: 
\[
\Fg = \lh \ba{c} 0 \\ v_1 + h/\sqrt{2} \ea \rh, \hs{2} \vf = v_2 + \frac{a}{\sqrt{2}}.
\]
Expanding the potential to quadratic order in the fields $h$ and $a$ leads to the scalar mass terms
\be
V = \frac{1}{2}\, \lb_1 v_1^2\, h^2 + \frac{1}{2}\, \lb_2 v_2^2\, a^2 + \frac{1}{2}\, \lb_m v_1 v_2\, ha + ...
\label{9.1}
\ee
This expression is diagonalised by substitution of the linear combinations
\be
h = h_+ \cos \thg_s - h_- \sin \thg_s, \hs{1} a = h_+ \sin \thg_s + h_- \cos \thg_s, 
\label{9.2}
\ee
where the scalar mixing angle is defined by
\be
\tan 2 \thg_s = \frac{\lb_m v_1 v_2}{\lb_1 v_1^2 - \lb_2 v_2^2} =
 \frac{\lb_m\, g_2 g_x\, m_Z m_{Z'} \cos \thg_w \cos \del}{\lb_1 g_x^2\, m_Z^2 \cos^2 \thg_w 
 - \lb_2 g_2^2\, m_{Z'}^2 \cos^2 \del}.
\label{9.3}
\ee
In terms of the mass eigenstates the potential becomes
\be
V = \frac{1}{2}\, m_+ h_+^2 + \frac{1}{2}\, m_-^2 h_-^2 + ...
\label{9.4}
\ee
with 
\be
\ba{lll}
m_{\pm} & = & \dsp{ \frac{1}{2} \left[ \lb_1 v_1^2 + \lb_2 v_2^2 
   \pm \sqrt{(\lb_1 v_1^2 - \lb_2 v_2^2)^2 + \lb_m^2 v_1^2 v_2^2} \right] }\\
  & & \\
  & = & \dsp{ \frac{1}{g_2^2 g_x^2} \left[ \lb_1 g_x^2\, m_Z^2 \cos^2 \thg_w + \lb_2 g_2^2\, m_{Z'}^2 \cos^2 \del \rd }\\
  & & \\
  & & \dsp{ \ld \pm\, \sqrt{ (\lb_1 g_x^2\, m_Z^2 \cos^2 \thg_w - \lb_2 g_2^2\, m_{Z'}^2 \cos^2 \del)^2
   + \lb_m^2 g_2^2 g_x^2\, m_Z^2 m_{Z'}^2 \cos^2 \thg_w \cos^2 \del} \right]. }  
\ea
\label{9.5}
\ee
In the limit $m_{Z'}^2 \gg m_Z^2$ this simplifies to
\be
\ba{l}
\dsp{ m_-^2 = m_H^2 = \frac{2 \lb_1 m_Z^2 \cos^2 \thg_w}{g_2^2} \lh 1 - \frac{\lb_m^2}{4\lb_1 \lb_2} + ...\rh, }\\
 \\
\dsp{ m_+^2 = \frac{2 \lb_2 m_{Z'}^2 \cos^2 \del}{g_x^2} + ... }
\ea
\label{9.6}
\ee
Thus in this limit the Higgs mass of lightest Higgs particle is lower than in the minimal Standard Model by the 
amount $\lb_m^2/4\lb_1 \lb_2$ (at tree level). Clearly the measurement of the Higgs mass $m_H$ does not 
suffice to compute the Higgs self-coupling $\lb_1$ unless $\lb_m^2 \ll \lb_1 \lb_2$.
\vs{2}

\nit
{\bf 11 Discussion}
\vs{1}

\nit
The Standard Model with right-handed neutrinos possesses an additional anomaly-free $U(1)$ symmetry, which 
can be gauged and gives rise to a new neutral vector boson $Z'$. In such an extension of the minimal Standard 
Model also the right-handed neutrinos participate in the gauge interactions; in this sense it completes the electro-weak 
gauge sector of the Standard Model. The minimal implementation of the scenario introduces a singlet Higgs scalar 
which can also provide a Majorana mass for neutrinos. The resulting theory has two free parameters, which may
be identified with $\xi$ and $\bg$ in eq.\ (\ref{7.r3}). It also introduces a new gauge coupling constant $g_x$, two 
new BEH couplings $(\lb_2, \lb_m)$ and the new BEH vacuum expectation value $v_2$ or equivalently the new 
vector-boson mass $m_{Z'}$.

In analyzing the theory I have focussed on determining the mass spectrum and the mass eigenstates. Their 
electroweak interactions are described in principle by the complete lagrangean as given in appendix A. However, 
as in the spontaneously broken phase of $SU_L(2) \times U_Y(1) \times U_X(1)$ almost all mass eigenstates are 
linear combinations of interaction eigenstates, the interactions of the mass eigenstates are not easy to read off 
directly. I leave this as an exercise for the reader, who can find keys to the solution in the literature listed in the
references.  

A number of new issues arise. First of all, the discussion in the preceding paragraphs were mostly phrased in 
terms of a single family of fermions. Family mixing was not discussed, but is easy to incorporate when all families 
have the same values of $\bg$. Although quarks and leptons in different families could a priori couple differently 
to the neutral vector bosons $Z$ and $Z'$, the observed quark mixing forbids this if the theory is to remain 
renormalizable and there is only one Higgs doublet. Of course the mixing matrices themselves affect different 
families in different ways, and this can also apply to the new Majorana-Yukawa terms for the right-handed 
neutrinos \ct{pontecorvo1957,mns1962}. 

The $Z'$ interactions can be incorporated in Grand Unified Theories in which $B-L$ is part of the gauge symmetry, 
such as $SO(10)$- or $E_6$-models, or a semi-GUT like $SU(5) \times U(1)$ \ct{ross2003}; also L-R-symmetric 
models with gauge group $SU_c(3) \times SU_L(2) \times SU_R(2) \times U_{B-L}(1)$ fit this scheme. Many of 
these models require specific values of $\xi$ and $\bg$. Supersymmetry is another feature that can be made 
consistent with the minimal extension of the Standard Model considered here, but also with the GUT-models in 
which it may be incorporated \ct{barger2009,sgn-jwvh1998,sgn-jwvh1999}.

A last issue to discuss is the mixing of $U(1)$ gauge bosons \ct{holdom1986,hook2011}. The 
$U_Y(1) \times U_X(1)$ invariance of the field strength tensors $F_{\mu\nu}(B)$ and $F_{\mu\nu}(C)$ 
allows a mixed kinetic term 
\be
- \frac{\kg}{2}\, F_{\mu\nu}(B) F^{\mu\nu}(C)
\label{11.1}
\ee
to appear in the action. Even if it does not appear at tree-level it may reappear in the effective quantum action 
as a result of renormalization; for example, fig.\ 3 shows the two gauge fields coupling to 
the same fermion loop. However, rediagonalizing the kinetic terms only adjusts the gauge couplings and 
resulting mixing angles $(\thg_w, \del)$ as a function of the renormalization scale, but will not change the 
basic physics involved at the TeV scale. At much higher energies the kinetic terms of the vector bosons may 
actually be protected by GUT-scale symmetries. For details one should consult more extensive reviews \ct{leike1998,langacker2009}.

\vs{-2}
\bc
\scalebox{0.14}{\includegraphics{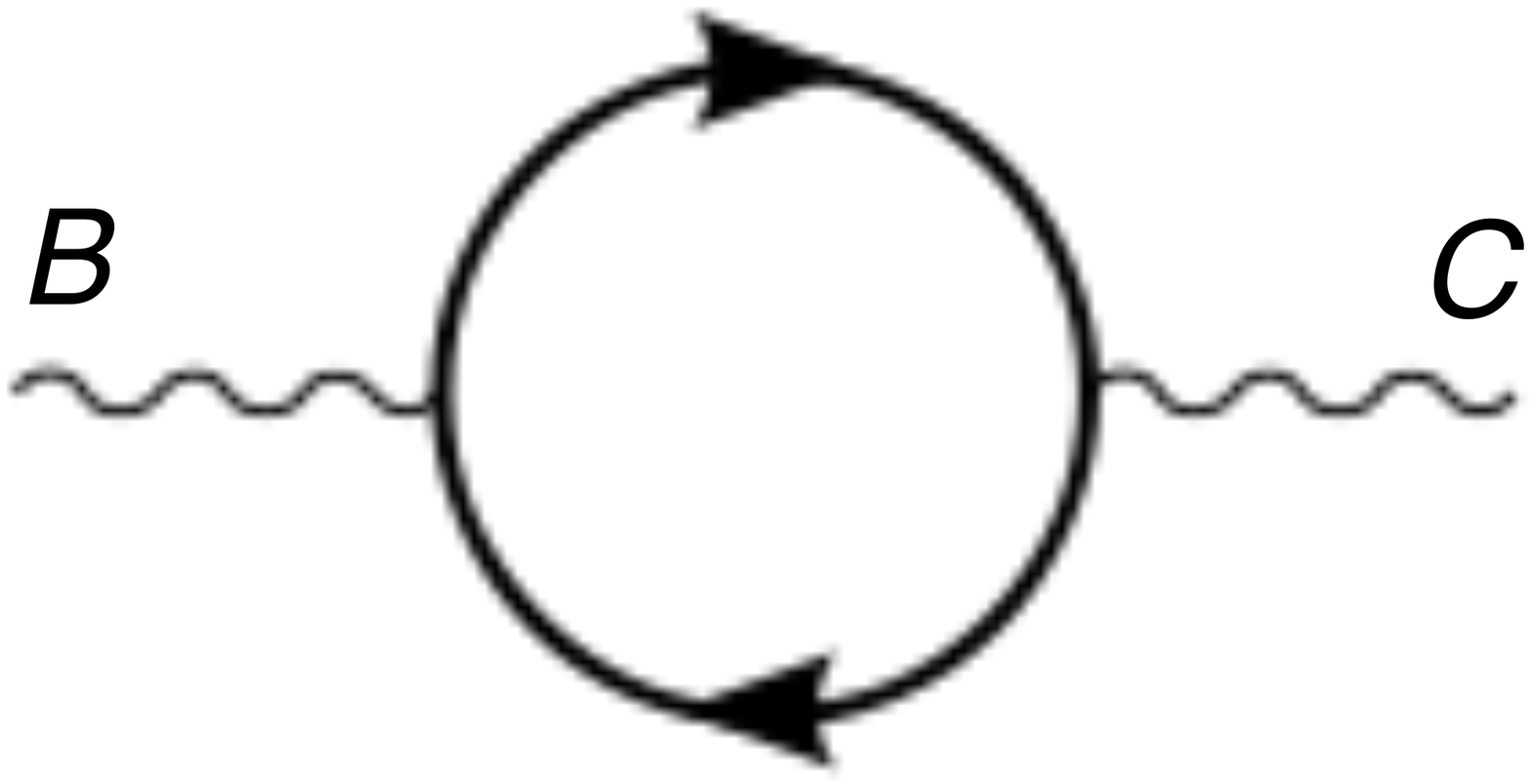}}
\vs{-3}

{\footnotesize Fig.\ 3: Fermion loop mixing $B$- and $C$-bosons.}
\ec

\np
\nit
{\bf Appendix} 
\vs{1}

\nit
{\bf A.\ Lagrangean} \\ 
Collecting all results for the $U_X(1)$-extended standard model the Lagrange density for the electroweak sector becomes
\be
\ba{lll}
\cL & = & \dsp{ - \frac{1}{4} \left[ F^i_{\mu\nu}(W)\right]^{2} - \frac{1}{4}\, F_{\mu\nu}(B)^2 - \frac{1}{4}\, F_{\mu\nu}(C)^2 
     - \left| D \Fg \right|^2 - | D\vf |^2 }\\
 & & \\
 & & \dsp{ +\, \frac{i}{2}\, \sum_L \bar{\Psi}_L \gam \cdot D \Psi_L + \frac{i}{2}\, \sum_R \bar{\Psi}_R \gam \cdot D \Psi_R }\\
 & & \\
 & & \dsp{ -\, \frac{\lb_1}{4} \lh |\Fg|^2 - v_1^2 \rh^2 - \frac{\lb_2}{4} \lh |\vf|^2 - v_2^2 \rh^2
    - \frac{\lb_m}{4} \lh |\vf|^2 - v_2^2 \rh \lh |\Fg|^2 - v_1^2 \rh }\\ 
 & & \\
 & & \dsp{ - \frac{i}{2}\, \eps_{ab} \Fg^a \left[ f_U \bar{U}_R Q^b_L + f_N \bar{N}_R L^b_L \right]  
                 - \frac{i}{2}\, \Fg^*_a \left[ f_D \bar{D}_R Q^a_L + f_E \bar{E}_R L^a_L \right] + c.c.\ }\\
 & & \\
 & & \dsp{ - \frac{1}{2} \lh f_x\, \vf^*\, N_R^T C N_R + f^*_x\, \vf\, \bar{N}_R C \bar{N}_R^T \rh. }                 
\ea
\label{L.1}
\ee 
In this expression
\be
\ba{l}
F_{\mu\nu}^i(W) = \der_{\mu}W^i_{\nu} - \der_{\nu} W^i_{\mu} - g_2 \eps^{ijk}\, W^j_{\mu} W^k_{\nu}, \\
 \\
F_{\mu\nu}(B) = \der_{\mu} B_{\nu} - \der_{\nu} B_{\mu}, \hs{1} 
F_{\mu\nu}(C) = \der_{\mu} C_{\nu} - \der_{\nu} C_{\mu};
\ea
\label{L.2}
\ee
and
\be
\ba{l}
D \Psi_L = \lh \der - \frac{i}{2} \lh g_x X C + g_1 Y B + g_2 \bfW \cdot \bftau \rh \rh \Psi_L,  \\
 \\
D \Psi_R = \lh \der - \frac{i}{2} \lh g_x X C + g_1 Y B \rh \rh \Psi_R.
\ea
\label{L.3}
\ee
Here the $\bftau$ refer to the triplet of Pauli matrices:
\[
\tau_1 = \lh \ba{cc} 0 & 1 \\ 1 & 0 \ea \rh, \hs{1} \tau_2 = \lh \ba{cc} 0 & -i \\ i & 0 \ea \rh, \hs{1}
\tau_3 = \lh \ba{cc} 1 & 0 \\ 0 & -1 \ea \rh.
\]
As a result 
\be
\bfW \cdot \bftau = W^1 \tau_1 + W^2 \tau_2 + W^3 \tau_3 = \lh \ba{cc} W^3 & W^1 - i W^2 \\  & \\ W^1 + i W^2 & -W^3 \ea \rh,
\label{L4}
\ee
and $\Psi_L$ couples to the vector-boson combinations
\be
g_x XC + g_1 YB + g_2 \bfW \cdot \bftau = \lh \ba{cc} g_x X C + g_1 Y B + g_2 W^3 & g_2 (W^1 - i W^2) \\
                                                                                           & \\
                                                                                       g_2( W^1 + i W^2) & g_x XC + g_1 YB - g_2 W^3 \ea \rh.
\label{L5}
\ee

\np
\nit
{\bf B.\ Notations and conventions} \\
The Minkowski metric is defined as 
\be
\eta_{\mu\nu} = \mbox{diag} (-1, +1, +1, +1).
\label{a.1}
\ee
The Dirac matrices $\gam_{\mu}$ satisfy
\be
\left\{ \gam_{\mu}, \gam_{\nu} \right\} = \gam_{\mu} \gam_{\nu} + \gam_{\nu} \gam_{\mu} 
= 2 \eta_{\mu\nu}\, {\bf 1}, 
\label{a.2}
\ee
and we take them such that 
\be
\gam_{\mu}^{\dagger} = \gam_0 \gam_{\mu} \gam_0. 
\label{a.3}
\ee
With the permutation tensor defined such that $\ve_{0123} = +1$ the chirality operator $\gam_5$ is
\be
\gam_5 = \frac{i}{4!}\, \ve^{\mu\nu\kg\lb} \gam_{\mu} \gam_{\nu} \gam_{\kg} \gam_{\lb}
 = - i \gam_0 \gam_1 \gam_2 \gam_3 \gam_4 = \gam_5^{\dagger}.
\label{a.4}
\ee
It has the anti-commutation property
\be
\left\{ \gam_5, \gam_{\mu} \right\} = 0.
\label{a.r1}
\ee
Another important operator is the charge-conjugation matrix $C = - C^T = C^{\dagger} = C^{-1}$ defined by
\be
C \gam_{\mu}^T = - \gam_{\mu} C .
\label{a.5}
\ee
A convenient representation is one in which $\gam_5$ is diagonal:
\be
\ba{l}
\dsp{ \gam_0 = \lh \ba{cc} 0 & i 1 \\
                                         i1 & 0 \ea \rh,  \hs{2} \gam_i = \lh \ba{cc} 0 & i \sg_i \\
                                                                                                  - i \sg_i & 0 \ea \rh, }\\
\\
\dsp{ \gam_5 = \lh \ba{cc} 1 & 0 \\
                                          0 & -1 \ea \rh, \hs{2} C = \lh \ba{cc} \sg_2 & 0 \\
                                                                                                   0 & -\sg_2 \ea \rh. }                                                                                                  
\ea
\label{a.6}
\ee
A Lorentz transformation on spinors $\psi$ with parameters $\og^{\mu\nu} = - \og_{\nu\mu}$ is realized by
\be
\psi' = e^{\frac{1}{2} \og^{\mu\nu} \sg_{\mu\nu}} \psi \approx 
\lh 1 + \frac{1}{2}\, \og^{\mu\nu} \sg_{\mu\nu} + ... \rh \psi, \hs{2}
\sg_{\mu\nu} = \frac{1}{4}\, \left[ \gam_{\mu}, \gam_{\nu} \right].
\label{a.7}
\ee
It follows directly from (\ref{a.r1}) that
\be
\left[ \gam_5, \sg_{\mu\nu} \right] = 0,
\label{a.8}
\ee
and the eigenvalues of $\gam_5$ are unchanged under Lorentz transformations. 

\np

\end{document}